\documentclass[aps,twocolumn]{revtex4}

\usepackage{amssymb}
\usepackage{graphicx}
\usepackage{amsmath}
\usepackage{epstopdf}
\usepackage{subfigure}

\begin{document}

\title{Long Trend Dynamics in Social Media}

\author{Chunyan Wang$^{1}$, Bernardo A. Huberman$^{2}$\footnote[1]{ Electronic address: bernardo.huberman@hp.com} }

\address{$^1$ Department of Applied Physics, Stanford University, CA, USA , $^2$ Social Computing Lab, HP Labs, Palo Alto, California, USA }


\begin{abstract}
A main characteristic of social media is that its diverse content, copiously generated by both standard  outlets and general users, constantly competes for the scarce attention of large audiences. Out of this flood of information some topics manage to get enough attention to become the most popular ones and thus to be prominently displayed as trends. Equally important, some of these trends persist long enough so as to shape part of the social agenda. How this happens is the focus of this paper. By introducing a stochastic dynamical model that takes into account the user's repeated involvement with given topics, we can predict the distribution of trend durations as well as the thresholds in popularity that lead to their emergence within social media. Detailed measurements of datasets from Twitter confirm the validity of the model and its predictions.   
\end{abstract}

\maketitle

\section{Introduction}
The past decade has witnessed an explosive growth of social media, creating a competitive environment where topics compete for the attention of users~\cite{M93,F08}. A main characteristic of social media is that both users and standard media outlets generate content at the same time in the form of news, videos and stories, leading to a flood of information from which it is hard for users to sort out the relevant pieces to concentrate on~\cite{E08,K10}. User attention is critical for the understand of how problems in culture, decision making and opinion formation evolve~\cite{Z92,W07,G05}. Several studies have shown that attention allocated to on-line content is distributed in a highly skewed fashion~\cite{B98,J00,B01,V06}. While most  documents receive a negligible amount of attention, a few items become extremely popular and persist as public trends for long a period of time~\cite{N90,K02,H11}. Recent studies have focused on the dynamical growth of attention on different kinds of social media, including Digg~\cite{F07, L10, J09}, Youtube~\cite{C08}, Wikipedia~\cite{J10, C06, Z06} and Twitter~\cite{B09}. The time-scale over which  content persists as a topic in these media also varies on a scale from hours to years. In the case of news and stories,  content spreads on the social network until its novelty decays~\cite{F07}. In information networks like Wikipedia, where a document remains alive for months and even years, popularity is governed by  bursts of sudden events and is explained by the rank shift model~\cite{J10}. 

While previous work has successfully addressed the growth and decay of news and topics in general, a remaining problem is  why some of the topics stay popular for longer periods of time than others and thus contribute to the social agenda. In this paper, we focus on  the dynamics of long trends and their persistence within social media. We first introduce a dynamic model of attention growth and derive the distribution of trend durations for all topics. By analyzing the resonating nature of the content within the community, we provide a threshold criterion that successfully predicts the long term persistence of social trends. The predictions of the model are then compared with measurements taken from Twitter, which as we show provides a validation of the proposed dynamics.

This paper is structured as follows. In Section 2 we describe our model for attention growth and the persistence of trends. Section 3 describes the data-set and the collection strategies used in the study, whereas Section 4 discusses the measurements made on  data-sets from Twitter and compares them with the predictions of the model. Section 5 concludes with a summary of our findings and future directions.

\section{Model}

On-line micro-blogging and social service websites enable  users to read and send text-based messages to certain topics of interest. The popularity of these topics is commonly measured by the number of postings about these topics~\cite{F07,J10}. For instance on Twitter, Digg and Youtube, users post their thoughts on topics of interest in the form of tweets and comments. One special characteristic of social media that has been ignored so far is that users can contribute to the popularity of a topic more than once. We take this into account by denoting first posts on a certain topic from a certain user by  the variable First Time Post, ($FTP$). If the same user posts on the topic more than once, we call it a Repeated Post, ($RP$). In what follows, we first look at the growth dynamics of $FTP$. 

When a topic first catches people's attention, a few people may further pass it on to others in the community. If we denote the cumulative number of $FTP$ mentioning the topic at time $t$ by $N_t$, the growth of attention can be described by $N_t = (1+\chi_t)N_{t-1}$, where the $\chi_t$ are assumed to be small, positive, independent and identically distributed random variables with mean $\mu$ and variance $\sigma^2$. For small $\chi_s$, the equation can be approximated as:
\begin{equation}
N_t \simeq \prod\limits_{s = 1}^t {e^{\chi_s}}N_0=e^{\sum\limits_{s
= 1}^t {\chi_s} }N_0.
\label{eq:Lognormal-N}
\end{equation}
Taking logarithms on both sides, we  obtain $\log \frac{{N_t }}{{N_0 }} = \sum\limits_{s=1}^t {\chi_s}$, Applying the central limit theorem to the sum, it follows that the cumulative count of $FTP$ should obey a log-normal distribution. 

We now consider the persistence of social trends. We use  the variable vitality, $\phi_t = \frac{N_t}{N_{t-1}}$, as a measurement of popularity, and assume that if the vitality of a topic falls below a certain threshold $\theta_1$, the topic stops trending. Thus
\begin{equation}
\log\phi_t = \log \frac{N_t}{N_{t-1}} =\log \frac{N_t}{N_{0}}-\log
\frac{N_{t-1}}{N_0} \simeq \chi_t.
\end{equation}
The probability of ceasing to trend at the time interval $s$ is equal to the probability that $\phi_s$ is lower than a threshold value $\theta_1$, which can be written as:
\begin{equation}
\begin{aligned}
p = \Pr(\phi_s<\theta_1)=\Pr(\log\phi_s<\log(\theta_1))\\
= \Pr(\chi_s<\log(\theta_1))= F(\log(\theta_1)),
\end{aligned}
\end{equation}
where $F(x)$ is the cumulative distribution function of the random variable $\chi$. We are thus able to determine the threshold value from $\theta_1  = e^{F ^{ - 1} (p)}$ if we know the distribution of the random variable $\chi$. Notice that if $\chi$ is independent and identically distributed, it follows that the distribution of trending durations is given by a geometric distribution with $\Pr(L=k)=(1-p)^{k}p$. The expected trending duration of a topic, $E(L)$, is therefore given by
\begin{equation}
E(L)  = \sum\limits_0^\infty {(1 - p)^k p \cdot k} = \frac{1}{p} -1=\frac{1}{F(\log(\theta_1))}-1.
\end{equation}

Thus far we have only considered the impact of $FTP$ on social trends by treating all topics as identical to each other. To account for the resonance between users and specific topics we now include the $RP$ into the dynamics. We define the instantaneous number of $FTP$ posted in the time interval $t$ as $FTP_t$, and the repeated posts, $RP$, in the time interval $t$ as $RP_t$. Similarly we denote the cumulative number of all posts-including both $FTP$ and $RP$-as $S_t$. The resonance level of fans with a given topic is measured by  $\mu_t=\frac{FTP_t+RP_t}{FTP_t}$, and we define the expected value of $\mu_t$, $E(\mu_t)$ as the active-ratio $a_q$. 

\begin{figure}[htl]
 \centering
 \includegraphics[width=7cm]{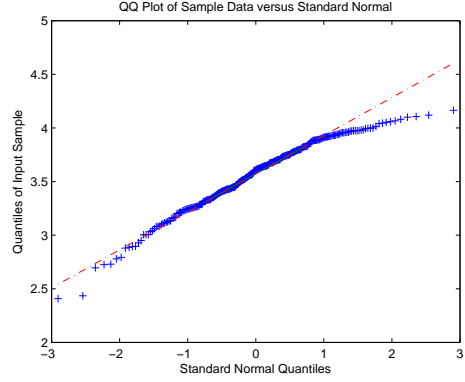}
 \caption{The normal Q-Q plot of $\log(N_{10})$. The straight line shows that the data follows a lognormal distribution with a slightly shorter tail.}
 \label{fig:QQPlot-N}
\end{figure}

We can simplify the dynamics by assuming that $\mu_t$ is independent and uniformly distributed on the interval $[1,2a_q-1]$. It then follows that the increment of $S_t$ is given by the sum of $FTP_t$ and $RP_t$. We thus have 
\begin{equation}
\begin{aligned}
S_t-S_{t-1} =FTP_t+RP_t = \mu_tFTP_t=\mu_t(N_t-N_{t-1})\\
=\mu_t\chi_tN_{t-1}.
\end{aligned}
\end{equation} 
And also
\begin{equation}
\begin{aligned}
E_\mu(S_t)=E_\mu(S_{t-1})+a_q(N_t-N_{t-1})=\\
E_\mu(S_{t-2})+a_q(N_{t}-N_{t-2})=\cdot\cdot\cdot=\\
E_\mu(S_0)+a_q(N_t-N_0) = a_qN_t.
\end{aligned}
\end{equation} 
We approximate $S_{t-1}$ by $\mu_tN_{t-1}$. Taking back to Eq. 5, we have 
\begin{equation}
S_t \simeq \mu_t(\chi_t+1)N_{t-1} \simeq \mu_te^{\chi_t}N_{t-1}.
\end{equation} 
From this, it follows that the dynamics of the full attention process is  determined by the two independent random variables, $\mu$ and $\chi$. Similarly to the derivation of  Eq. 3, the topic is assumed to stop trending if the value of either one of the random variables governing the process falls below the thresholds $\theta_1$ and $\theta_2$, respectively. The probability of ceasing to trend, defined as $p^\star$, is now given by  
\begin{equation}
p^\star = \Pr(\chi_t<\log(\theta_1))\Pr(\mu_t<\theta_2)=\frac{\theta_2-1}{2(a_q-1)}p,
\end{equation} 
$p = F(\log(\theta_1))$. The expected value of $L_q$ for any topic $q$ is given by
\begin{equation}
E(L_q)=\frac{2(a_q-1)}{F(\log{\theta_1})(\theta_2-1)}-1.
\end{equation} 
Which states that the persistent duration of trends associated with given topics is expected to scale linearly with the topic users' active-ratio. From this result it follows that one can  predict the trend duration for any topic by measuring its user active-ratio after the values of $\theta_1$ and $\theta_2$ are determined from empirical observations.

\begin{figure}[htl]
  \centering
  \subfigure[Frequency Plot] {\includegraphics[height=5.0cm,width=6.0cm]{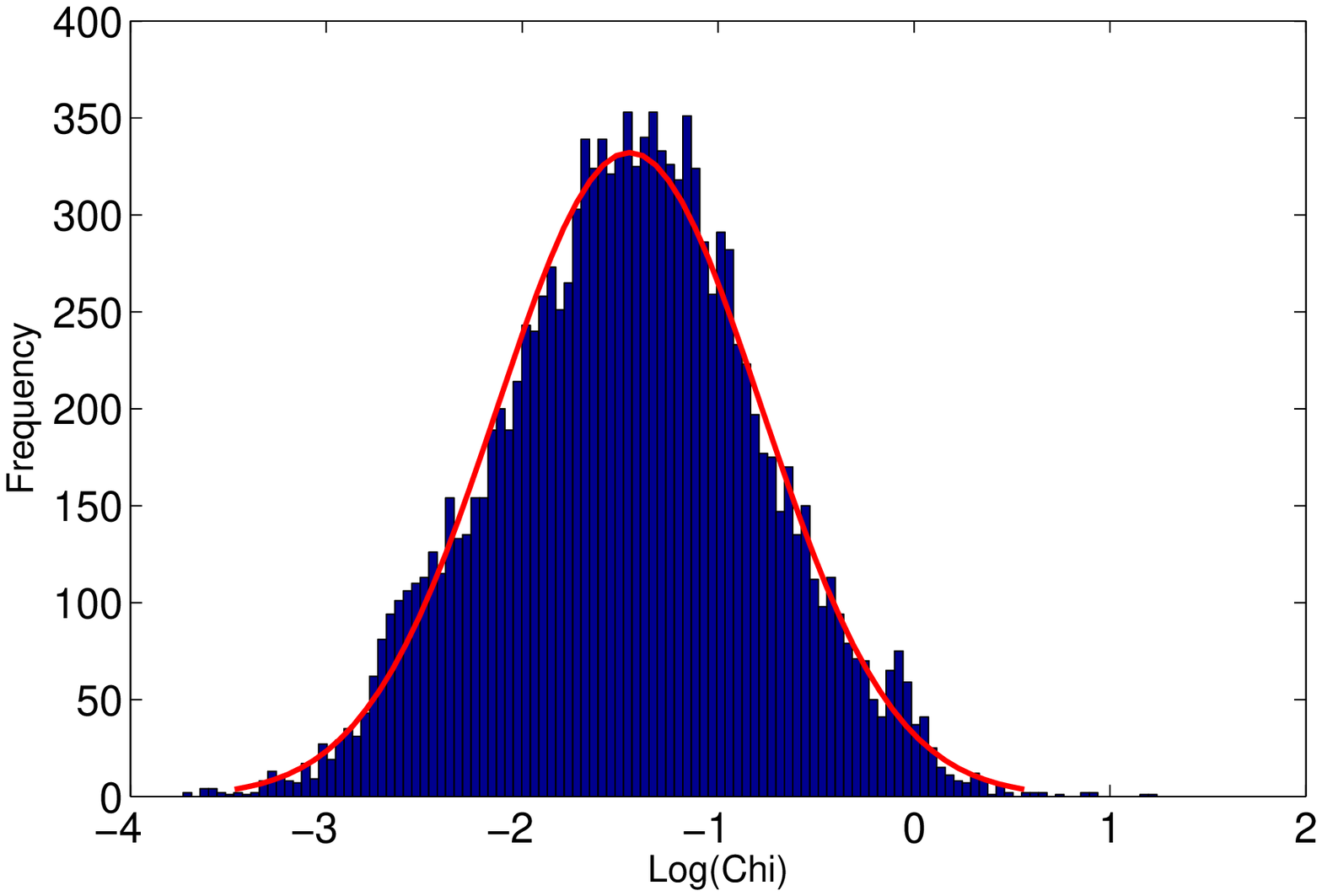}} 
  \subfigure[Q-Q Plot]{\includegraphics[height=5.0cm,width=6.0cm]{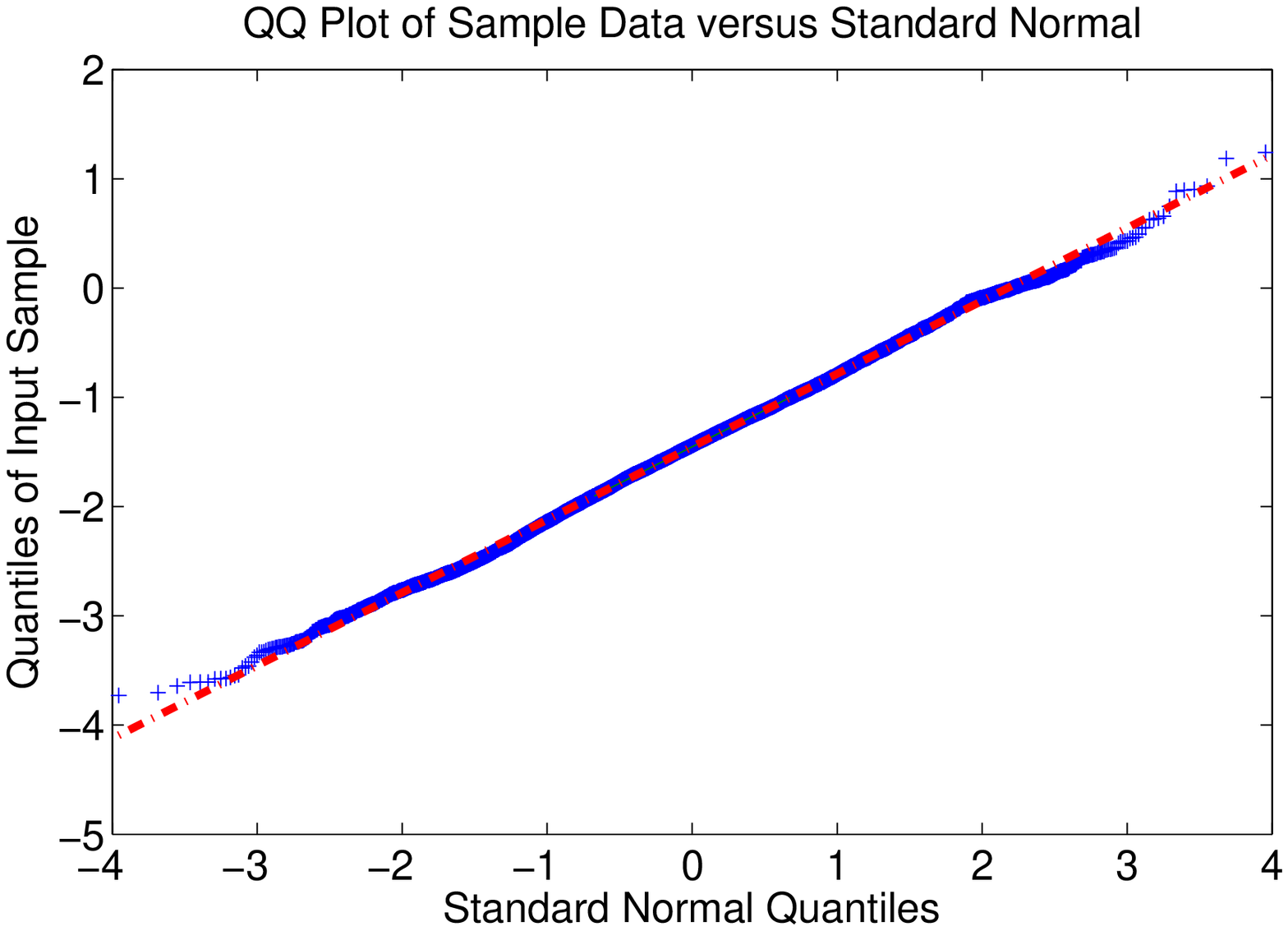}}
 \caption{Distribution of $\chi$ over different $t$ and all social trends. $\log(\chi)$ is following a normal distribution with mean equal to $-1.4522$ and standard deviation equal to $0.6715$. (a) The frequency plot of $\log(\chi)$. (b) The Q-Q plot of $\log(\chi)$}
 \label{fig:Chifrequency}
\end{figure}

\section{DATA}
To test the  predictions of our dynamic model, we analyzed data from Twitter, an extremely popular social network website used by over 200 million users around the world. Its interface of allows users to post short messages, known as tweets, that can be read and retweeted by other Twitter users. Users declare the people they follow, and they get notified when there is a new post from any of these people. A user can also forward the original post of another user to his followers by the re-tweet mechanism. 

In our study, the cumulative count of tweets and re-tweets that are related to a certain topic was used as a proxy for the popularity of the topic. On the front page of Twitter there is also a column named trends that presents the few keywords or sentences that are most frequently mentioned in Twitter at a given moment. The list of popular topics in the trends column is updated every few minutes as new topics become popular. We collected the topics in the trends column by performing an API query every 20 minutes. For each of the topics in the trending column, we used the Search API function to collect the full list of tweets and re-tweets  related to the topic over the past 20 minutes. We also collected information about the author of the post, identified by a unique user-id, the text of the post and the time of its posting. We thus obtained a dataset of 16.32 million posts on 3361 different topics. The longest trending topic we observed had a length of $14.7$ days. We found that of all the posts in our dataset, $17\%$ belonged to the $RP$ category. 

\begin{figure}[htl]
 \centering
 \includegraphics[width=7.0cm]{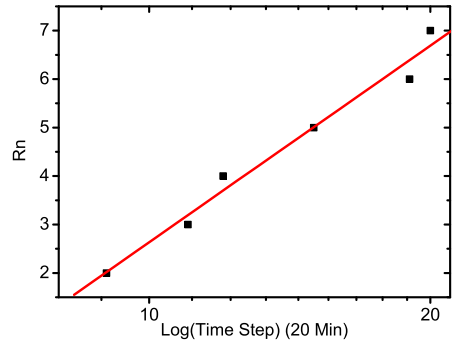}
 \caption{The linear scaling relationship between $R_n$ and $\log(t)$ of topic 'Kim Chul Hee', a Korean pop star. The topic kept trending for 14 days on Twitter in September 2010. The number of records that have occurred up to time $t$ scales linearly with $\log(t)$.}
 \label{fig:Recordbreaking}
\end{figure}

\section{Results}

From the data-set that we analyzed, we found out that for a fixed time interval of 200 minutes at $t=10$ $N_t$ all trends follows a log-normal distribution. As can be seen from Figure~\ref{fig:QQPlot-N}, the normal Q-Q plot of $\log(N_{10})$ follows a straight line. Different values of $t$ yield similar results. The Kolmogorov-Smirnov normality test of $\log(N_{10})$ with mean $3.5577$ and standard deviation $0.3266$ yields a P-value of $0.0838$. At a significance level of $0.05$, the test fails to reject the null hypothesis that $\log(N_{10})$ follows normal distribution, a result which is consistent with Equation~\ref{eq:Lognormal-N}. 

\begin{figure}[htl]
 \centering
 \includegraphics[width=7.0cm]{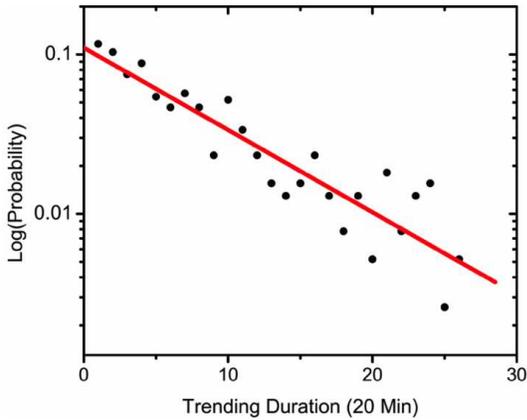}
 \caption{Semi-log plot of trending duration density. The straight line suggests an exponential family of the trending time distribution. The red line gives a fitting with R-square $0.9112$.}
 \label{fig:Durationdist}
\end{figure}

\begin{figure}[htl]
 \centering
 \includegraphics[width=7.0cm]{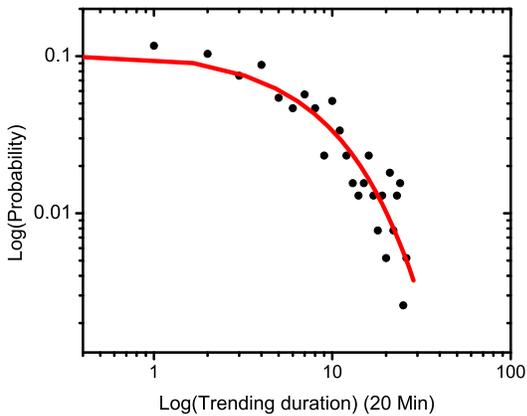}
 \caption{Density plot of trending duration in log-log scale. The distribution of duration deviates from a power law.}
 \label{fig:Durationdist2}
\end{figure}

We also observed that the distribution of $\chi$ from $\chi_t = \frac{N_t}{N_{t-1}}-1$. $\log(\chi)$ follows a normal distribution with mean equal to $-1.4522$ and a standard deviation value of $0.6715$, as shown in Figure~\ref{fig:Chifrequency}. The Kolmogorov-Smirnov normality test statistic gives a high p-value of $0.5346$. The mean value of $\chi$ is $0.0353$, which is small for the approximations in Equation~\ref{eq:Lognormal-N} and Equation~\ref{eq:} to be valid. We also examined the record breaking values of vitality, $\phi_t = \chi_t + 1$, which signal the behavior of the longest lasting trends. From the theory of records, if the values $\phi_t$ come from an independent and identical distribution, the number of  records that have occurred up to time t, defined as $R_n$, should scale linearly with $\log(t)$~\cite{R06,K07}. As is customary, we say that a new record has been established if the vitality of the trend at the moment is longer than all of the previous observations. As can be seen from Figure~\ref{fig:Recordbreaking}, this linear scaling is observed over a wide variety of topics. One implication of this observation is that confirms the validity of our assumption that the values of $\chi_1, \chi_2 ,\cdot\cdot\cdot, \chi_t$ are independent and identically distributed. 

\begin{figure}[tbp]
 \centering
 \includegraphics[height=6cm,width=7.0cm]{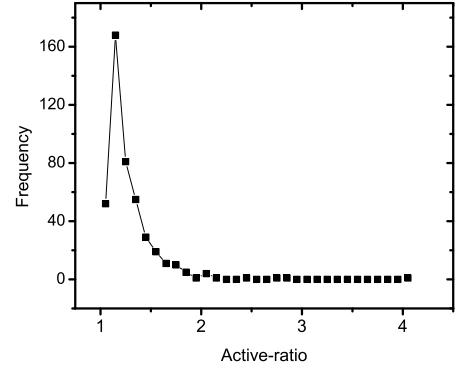}
 \caption{Frequency count of active-ratio over all topics. The maximum  ratio is $1.2$ among all topics.}
 \label{fig:Activeratiodist}
\end{figure}

Next we turn our attention to the distribution of durations of long trends. As shown in Figure~\ref{fig:Durationdist} and Figure~\ref{fig:Durationdist2}, a linear fit of trend duration as a function of density in a logarithmic scale suggests an exponential family, which is consistent with Eq. 4. The R-square of the fit has a value of 0.9112. From the log-log scale plot in Figure~\ref{fig:Durationdist2}, we observe that the distribution deviates from a power law, which is a characteristic of social trends that originate from news on social media~\cite{S10}. From the distribution of trending times, $p$ is estimated to have a value of $0.12$. Together with the measured distribution of $\chi$ and  Eq. 3, we can estimate the value of $\theta$ to be $1.0132$.  

\begin{figure}[htl]
 \centering
 \includegraphics[width=7.0cm,height=7cm]{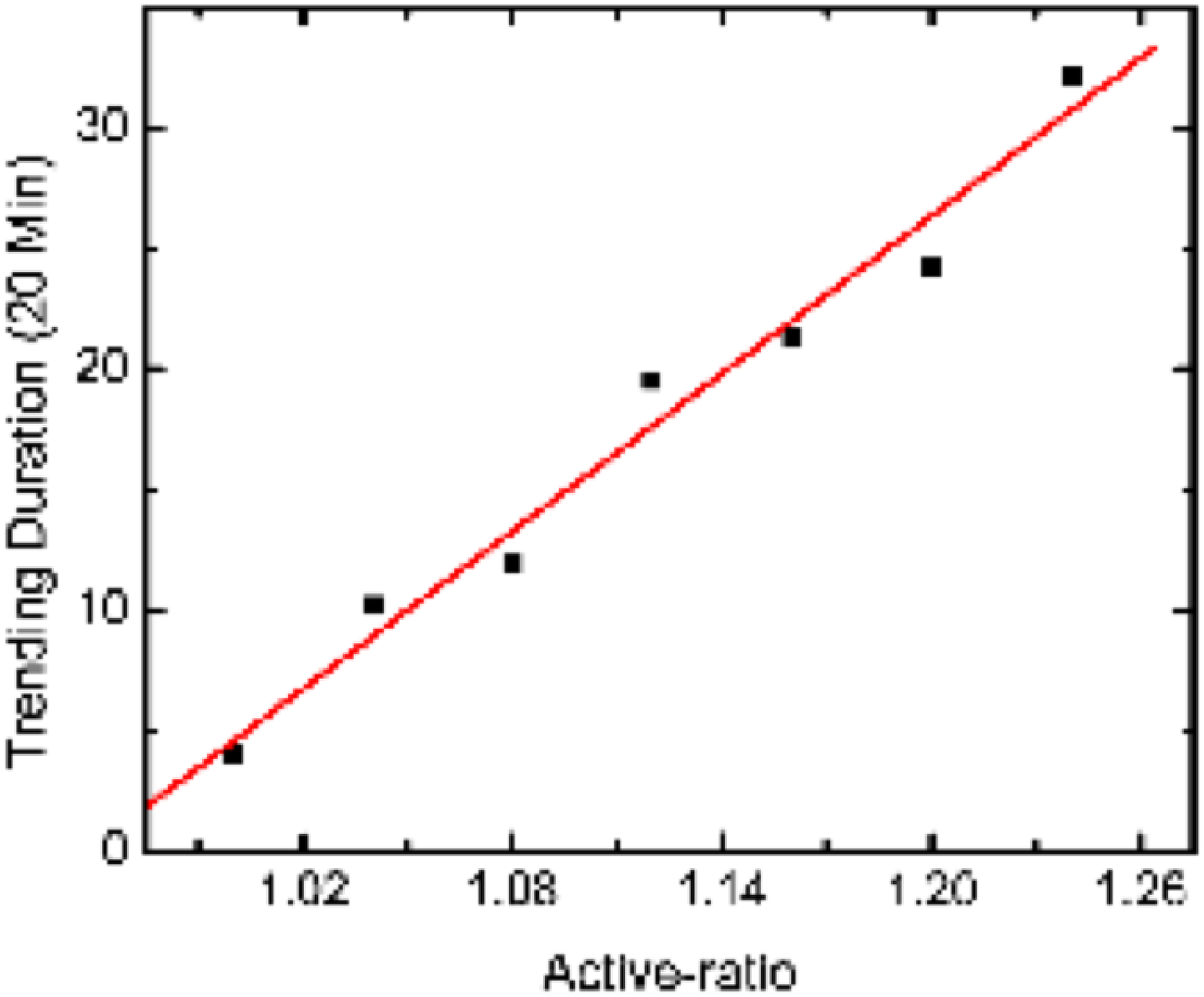}
 \caption{Linear relationship between trending duration and active-ratio, in good agreement with the predictions of model. }
 \label{fig:Predicttrending}
\end{figure}

We can also determine the expected duration of trend times stemming from the impact of active-ratio. The frequency count of active-ratios over different topics is shown in Figure~\ref{fig:Activeratiodist}, with a peak at $a_q = 1.2$. As can be seen in Figure~\ref{fig:Predicttrending}, the trend duration of different topics scales linearly with the active-ratio, which is consistent with the prediction of Eq. 9. The R-square of the linear fitting has a value of $0.98664$. From the slope of the linear fit and $\theta_1=1.0132$, and Eq. 9 we obtain a value for $\theta_2=1.153$. With the value of $\theta_1$ and $\theta_2$, we are able to predict the expected trend duration of any given topic based on measurements of its active-ratio.

\section{Discussion and Conclusion}
In this paper we investigated the persistence dynamics of  trends in social media. By introducing a stochastic dynamic model that takes into account the user's repeated involvement with given topics, we are able to predict the distribution of trend durations as well as the thresholds in popularity that lead to the emergence of given topics as trends within social media. The predictions of our mode were confirmed by a careful analysis of a data from Twitter. Furthermore, a linear relationship between the resonance level of users with given topics, and the trending duration of a topic was derived. The predictive power of this model provides a deeper understanding the popularity of on-line contents. Possible refinements may include the effect of competition between topics, sudden burst of events, the effect of  marketing campaigns, or any combination of them. In closing, we note that although the focus in this paper has been on trend dynamics that are featured on social media websites, the framework and model may be suitable to other types of content and off-line trends. The issue raised - that is, trending phenomenon under the impact of user's repeated involvement - is therefore a general one and should provide ample opportunities for future work. 

\section{Acknowledgments} We acknowledge useful discussions with S. Asur and G. Szabo. C.W. would like to thank HP Labs  financial support.

\bibliographystyle{abbrv}

\end{document}